\documentclass[submission,copyright,creativecommons]{eptcs}
\providecommand{\event}{THedu'11 Post-Proceedings}
\usepackage[utf8x]{inputenc}

\usepackage{url}
\usepackage{graphicx}

\newcommand{\webgeometrylab}{{\em Web Geometry Laboratory}}

\title{Integrating DGSs and GATPs in an Adaptative and Collaborative
  Blended-Learning Web-Environment}

\author{Vanda Santos
\institute{ESTGV -- IPV\\ Viseu, Portugal}
\institute{CISUC, University of Coimbra\\ Coimbra, Portugal}
\email{vsantos@estv.ipv.pt}
\and		
Pedro Quaresma
\institute{CISUC/Department of Mathematics\\ University of Coimbra\\
  3001-454 Coimbra, Portugal}
\email{pedro@mat.uc.pt}}

\def\titlerunning{Integrating DGSs and GATPs in an Adaptative
  Blended-Learning Web-Environment} \def\authorrunning{Santos, Vanda
  \& Quaresma.  Pedro}

\begin{document}
\maketitle

 \begin{abstract}
   The area of geometry with its very strong and appealing visual
   contents and its also strong and appealing connection between the
   visual content and its formal specification, is an area where
   computational tools can enhance, in a significant way, the learning
   environments.

   The common experience is that dynamic geometry software systems
   (DGSs) significantly help students to acquire knowledge about
   geometric objects and, more generally, to acquire mathematical
   rigour. Using geometry automated theorem provers (GATPs) capable of
   construction validation and production of human readable proofs,
   will consolidate the knowledge acquired with the use of the DGSs.
   It will also help studentes to comprehend the connection between a
   concrete representation of a particular geometric construction and
   its formal description.

   An adaptative and collaborative blended-learning Web-environment
   where the DGS and GATP features could be fully explored would be,
   in our opinion, a very rich and challenging learning environment for
   teachers and students.

   In this text we will describe the Web Geometry Laboratory, a Web
   environment incorporating a DGS and a repository of geometric
   problems, that can be used in a synchronous and asynchronous
   fashion and with some adaptative and collaborative features.
  

   As future work we want to enhance the adaptative and collaborative
   aspects of the environment and also to incorporate a GATP,
   constructing a learning environment for geometry, a dynamic system
   where the student has at his/her disposal the necessary tools for
   the study of theories and models of geometry. A system in which 
   students can be challenged by new problems. A system that allows
   students the opportunity to collaboratively develop and improve
   their study in the area of geometry anywhere, at anytime and at
   their own pace.
 \end{abstract}

\section{Introduction}
\label{sec:introduction}

The use of computational tools in a learning environment can greatly
enhance its dynamic, adaptative and collaborative features. It could
also extend the learning environment from the classroom to outside of
the fixed walls of the school. In the high-school curriculum in Portugal
the use of such tools is praised. Quoting from the official curriculum
specification\footnote{Translated from the Portuguese
  text.}~\cite{MatematicaA2011}:

\begin{quotation}
  The computer, by its own potential, namely in the areas of dynamic
  geometry, function representation and simulation, allows activities
  of exploration and research and also activities of recovery and
  development, in such a way that it constitutes an important asset to
  teachers and students, its use should be considered compulsory in
  this curriculum.
\end{quotation}

The dynamic geometry software systems (DGSs) allow an easy
construction of geometric figures built from free objects, elementary
constructions and constructed objects.  The dynamic nature of such
tools allows its users to manipulate the positions of the free objects
in such a way that the constructed objects are also changed, yet
preserving the geometric properties of the construction. These
manipulations are not formal proofs, the user is considering only a
finite set of concrete positions, nevertheless these tools provide a
first, not yet formal, link between the theories and models of
geometry.

The Pythagoras theorem is a good example where a DGS can be used in
a very fruitful way. The dynamic component allows us to work the
visuals proofs of this theorem. The JGEX Web page~\cite{Chou2011} has
some very nice examples of visual proofs of this theorem. 

The DGSs allow also to perform more complex geometric transformations
like translations, reflections, rotations. The advantages of the DGSs
in a learning environment are multiple: they are easy to use, they
stimulate the creativity and the discovery process. There are multiple
DGS available: GeoGebra, Cinderella, GeometerSketchpad, C.a.R., Cabri
and
GCLC~\cite{CaR2011,hohenwarter2002,Jackiw2001,Janicic06c,Laborde90,Richter-Gebert99}
to name some of the most used.

Automated theorem provers are less widely used as tools in a learning
environment, but geometry with its axiomatic nature is a ``natural''
area for a formal tool such as the GATP.  The GATPs give its users the
possibility to reason about a given DGS construction, this is no
longer a ``proof by testing'', but an actual formal proof.  If the
GATP is capable of producing synthetic proofs, the proof itself can be
an object of study, in other cases only the conclusion
matters~\cite{Chou96,Janicic06a}.  Another link between the GATPs and
the DGSs is given by the automated deductive testing, by the GATP, of
the soundness of the constructions made by the DGS~\cite{Janicic06b}.
Most, if not all, DGSs are capable of detecting and report syntactic
and semantic errors, but the verification of the soundness of the
construction is beyond their capabilities. If we have this kind of
integration between DGSs and GATPs we can check the soundness of a
given construction. Again, if the GATP produces synthetic proofs, the
proof itself can be an object of study, providing a logical
explanation for the error in the construction.  In either cases
(formal proofs or soundness of the constructions) we claim that the
GATPs can be used in the learning process~\cite{Janicic06b}.

To build an adaptative and collaborative blended-learning environment
for geometry, we claim that we should integrate a DGS, a GATP, a
repository of geometric problems (RGP), and the student model of
interaction\footnote{These last two with the help of a database
  management system (DBMS)} in a Web system capable of asynchronous
and synchronous interactions. A system with that level of integration
will allow building an environment where each student can have a broad
experimental, but with a strong formal support, learning platform.

Such an integration is still to be done, there are already many
excellent DGSs, some of them have some sort of integration with GATPs,
others with RGPs~\cite{Janicic06a,Quaresma06d}. Some attempts to
integrate these tools in a learning management system (LMS) have
already been done~\cite{Santos08}, but, as far as we know, all these
integrations are only partial integrations. A learning environment
where all these tools are integrated and can be used in a fruitful
fashion does not exist yet.

In this text we describe the Web Geometry Laboratory (WGL) an
asynchronous/synchronous Web environment that integrates a DGS program
and a repository of geometric problems, aiming to provide an
adaptative and collaborative blended-learning environment for
geometry.

\paragraph{\bf Paper overview.} In
section~\ref{sec:AdaptativeBlendedLearningWebEnvironmentGeometry} we
will describe the features needed in an adaptative and collaborative
blended-learning Web-environment for geometry. In
section~\ref{sec:WebGeometryLab} we present the Web Geometry
Laboratory, a system that aims to be an environment of that type, we
also describe the technical challenges that had to be tackled in its
implementation. In section~\ref{sec:CaseStudy} we present a case study
and in section~\ref{sec:ConclusionsFutureWork} we speak about the work
still to be done and we draw some final conclusions.


\section{Goals for an Adaptative and Collaborative Blended-Learning
  Web-En\-vi\-ro\-nment for Geometry}
\label{sec:AdaptativeBlendedLearningWebEnvironmentGeometry}

If we want to build an adaptative and collaborative blended-learning
environment for geometry what are the features and tools we are looking
for? The following list is, in our opinion, close to a complete,
minimal, set of the features and tools needed for building such a system.

\subsection{Geometric Tools}
\label{sec:geometrictools}

As said above, the advantages of the DGSs in a learning environment are
multiple: they are easy to use, they stimulate the creativity and the
discovery process. They provide an outstanding tool to substitute the
old ruler and compass used in the classrooms. The constructions made
from free objects and constructed objects allow a degree of property
preserving manipulations much superior to the capabilities of
physical tools.

In spite of the DGSs outstanding features, they do not create a
learning environment by themselves, so its integration in a learning
environment will be beneficial. The DGSs excel in the dynamic
construction of geometric figures, the learning environment should add
to this, the collaborative and adaptative features.

The DGSs provide a first, not yet formal, link between the theories
and models of geometry. But if we want to reason about the
constructions we are doing, to make conjectures about their
proprieties or in a more generic way to make formal deductive
reasoning about geometric constructions, we need more then a DGS, we
need to extend the reasoning from concrete instances in a given model
to formal deductive reasoning in a geometric theory.  To have this we
need to add to our environment a GATP capable of synthetic proofs
(e.g. the Area Method~\cite{Janicic2010}) or algebraic proofs (e.g.
the Wu's Method~\cite{Wu00} or the Gr\"obner Basis
Method~\cite{Kapur86}).

A GATP will be helpful, at least, at two distinct tasks: 

\begin{itemize}
\item Construction validation: most (if not all) DGSs use only
  geometric concepts interpreted from concrete instances in the
  Cartesian plane. A construction is always associated with a concrete
  fixed set of points (with concrete Cartesian coordinates). In such
  environments, some constructions are illegal (e.g., if they attempt
  to make the intersection of parallel lines), but the question if
  such construction is always illegal or it is illegal only for a
  given particular set of fixed points is left open.  Indeed, for
  answering such questions, one has to use deductive reasoning, and
  not only a semantic check for that special case. On those situations
  using a GATP we could get the information that the construction is
  illegal, and moreover, that it is illegal not only for a given
  special case, but always.  In this way, the deductive nature of
  geometrical conjectures and proofs are linked to the semantic
  nature of models of geometry and, also, to human intuition and to
  geometric visualisations~\cite{Janicic06b}.

\item Geometric proofs: proofs are exemplary mathematical
  contents. They can serve in mathematical education, aimed at
  acquiring mathematical rigour.

  Using a synthetic proof style GATP, e.g. an area method based
  GATP~\cite{Janicic2010}, we may have access to the proof itself. In
  these cases the proof itself will be an object of study.

\end{itemize}

An integration between a DGS and a GATP like the integration that can
be found in the GCLC system~\cite{Janicic06c,Janicic06a} provides an
environment where a user can reason about geometric conjectures,
learning about the links between a formal theory and its models.

The connection with a repository of geometric problems (RGPs) will add
to the environment, memory, i.e., the capability to save/recover
geometric constructions. The teacher would be able to prepare in
advance a set of constructions/exercises to be released to the
students. The students would be able to keep their constructions in a
personal {\em scrapbook}.

With a connection to a database management system (DBMS) we could
foresee the possibility of having a repository of geometric
constructions and/or individualised scrapbooks, but also the
possibility of constructing the students' model, keeping the history
of the students' interaction with the system, in this way allowing the
adaptation of the environment to each student~\cite{Iglezakisd04}.

\subsection{Blended-Environment}
\label{sec:blendedenvironment}

A learning environment for geometry should be, in our opinion, a
blended-learning environment\footnote{A blended-learning environment
  is a mixing of different learning environments. It combines
  traditional face-to-face classroom (synchronous) methods with more
  modern computer-mediated (asynchronous) activities.}. It should be
an environment that can be used as a geometry laboratory in a
classroom by teachers and students in a much enhanced substitute for
the ruler and compass physical instruments. But it should also allow
to extend itself outside the classroom, for homework tasks, for
problems proposed by the teacher to be solved outside the classroom,
for students tutored study, where the tutors could be the teacher(s)
and/or classmates.

A Web-environment is appropriate for both situations, in a classroom a
local or wide area network (LAN or WAN) environment would allow the
synchronous interaction between teacher and students. Outside the
classroom a WAN environment would allow synchronous and asynchronous
interactions. The ubiquity of the Web nowadays make such a learning
environment easily accessible.

\subsection{Collaborative Environment}
\label{sec:collaborativeEnvironment}

The environment should be collaborative, i.e. it should allow the
knowledge to emerge and appear through interaction between its
users~\cite{Inoue10}. The teacher and students should be able to
interact with each other but the environment should also be open to
interaction between students. Speaking about a geometric environment
this interaction should not be restricted to textual contents, (chats,
wikis, etc.) but it should be extended to geometric contents. i.e. the
users of such a system should be able to exchange geometric
constructions, or even to build a geometric construction in a
collaborative way. This exchange of geometric contents will open new
possibilities in terms of a collaborative geometric learning
environment.

\subsection{Adaptative Environment}
\label{sec:adaptativeEnvironment}

The environment should be adaptative, i.e. it should adapt the help
information given to different users and also, an important feature in
a learning environment, to adapt the learning path to the different
users needs~\cite{Iglezakisd04,Mora2001,Moriyon2008}.

The system should be able to infer the geometric knowledge of the
users or to use plan recognition, in terms of geometric knowledge, to
infer the actual plan or information goal of the
users~\cite{Iglezakisd04}. We see this as an important challenge to be
addressed by any new system.


Defining a network environment that integrates a DGS, a GATP and a
DBMS (for the RGP and for the students' model) will provide more
then the simple sum of its components. It will allow the construction
of an adaptative and collaborative blended-learning environment for
geometry. 

\section{Web Geometry Laboratory}
\label{sec:WebGeometryLab}

\begin{figure}[hbtp]
  \centering
  \includegraphics[scale=0.55]{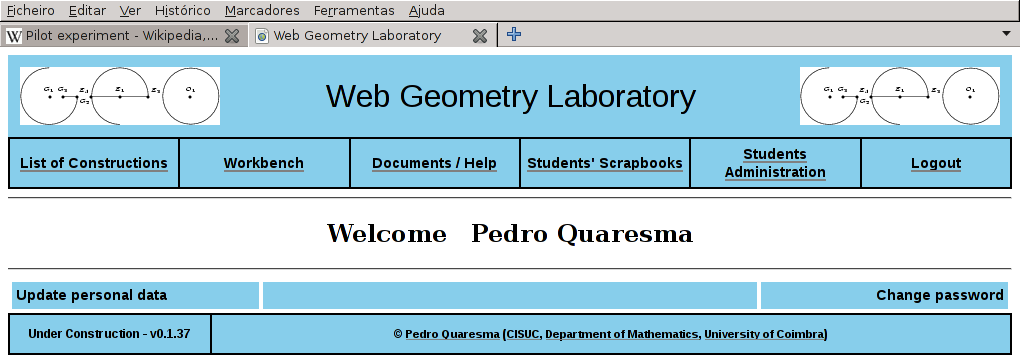}
  
  \caption{Web Geometry Laboratory --- Teacher's Login}
  \label{fig:wglTeacherLogin}
\end{figure}

\begin{figure}[hbtp]
  \centering
  \includegraphics[scale=0.355]{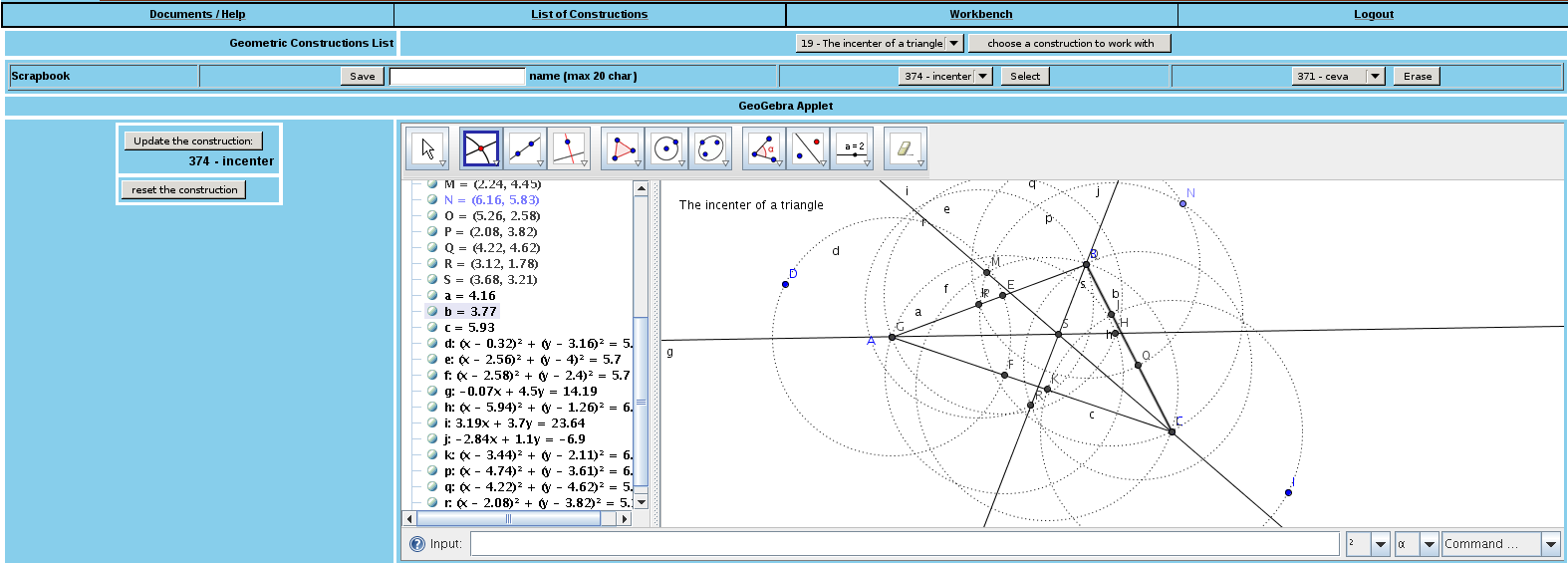}
  
  \caption{Web Geometry Laboratory --- Workbench}
  \label{fig:wglWorkbench}
\end{figure}

The {\em Web Geometry Laboratory}\footnote{A test site is available at
  \url{http://hilbert.mat.uc.pt/WebGeometryLab/}} (WGL) is a Web
environment (see Figures~\ref{fig:wglTeacherLogin}
and~\ref{fig:wglWorkbench}), that integrates a DGS program and a
repository of geometric problems (RGP), aiming to provide an
adaptative and collaborative blended-learning environment for
geometry.

The WGL system allows the teacher to create, store and provide a set
of geometric constructions to its students. The WGL system allows the
student to access the professor's constructions as well as those kept
in a personal {\em scrapbook}. The students' scrapbook (see
Figure~\ref{fig:wglWorkbench}) is a place where the student will keep
his/her own constructions, solutions of problems placed by the teacher
and/or his/her own exploratory activities. The teacher has also access
to the constructions made, or being made, by the students as a way to
be able to help the student during a class or to evaluate the work
done after class, or even as a mean to broadcast the work done by a
student to the rest of the class. In a next version of WGL the
students will be able to work collaboratively, seeing and exchanging
each other constructions.


The WGL aims to provide a learning environment for geometry using all
the potential of a given DGS, all the easiness of access provided by an
Web platform and with an individualised memory provided by a database
where all the history of each student is kept. With the inclusion of a
GATP and the implementation of some, now missing, features, the WGL
aims to became an adaptative and collaborative blended-learning
environment for geometry.

From the point of view of the server the system aims to be easy to
install, maintain and use. From the point of view of the clients,
teachers and students, the only feature that is needed is the access
to a Java aware\footnote{The WGL uses the GeoGebra Applet} Web
browser, which is, nowadays, a minor issue given the ubiquity of such
type of program.

If installed in the school server the WGL will not be confined to the
classroom, it can be opened to the global network, thus enabling its
remote use by teachers and students, extending its use to a
blended-learning environment. The WGL will allow teachers to set new
challenges to be solved outside the classroom, allowing each
individual student to work at his/her own pace.

Defining a network environment that integrates a DGS, a GATP and a
database, the {\webgeometrylab} will provide more then the simple sum
of its components. Integrating the DGS and the GATP fully in a
Web-environment will open the use of the DGS and GATP to a
collaborative blended-learning environment. Linking all that with a
database will allow an individualisation of the learning environment
with the creation of an adaptative (individualised) learning
environment.

\subsection{Levels of Access}
\label{sec:wglModules}

The access and privileges of the WGL mimics the Unix access and
privileges system. Any user of the WGL will belong to one of the three
possible levels of access, giving him/her a different initial Web-page
and a different set of privileges over the geometric constructions on
the repository.

As said above, the access to WGL has an hierarchy with three levels of
access: administrators, teachers and students.

\begin{itemize}
\item the {\em administrator's level\/} that allows the system
  administrator to define whose users will have {\em teachers
    status\/} within the system;
\item the {\em teachers' level\/} (notice the menu bar in
  Figure~\ref{fig:wglTeacherLogin}), which allows access (with
  validation) for teachers. Through this module it will be possible to
  teachers to make the management of students. It provides also access
  to the workbench in order to build and set the constructions that
  will be accessible to all the students, and finally it gives access
  to the students' {\em scrapbooks\/};
\item the {\em students' level\/} provides access, through a
  validation process, for students (see Figure~\ref{fig:wglWorkbench},
  notice the menu options). This module allows access, read-only, to
  the constructions provided by the teachers. Full access to the
  student's personal {\em scrapbook}. Last (but not least) each
  student will be able to open his/her working area (read and write
  access) to others in order to be able to work collaboratively.
\end{itemize}

In the actual version of WGL, apart from the definition of the
different levels of access, the definition of the privileges has a
very naive implementation. Any construction in the repository has a
``level attribute''. Negative numbers in this attribute means that the
construction is only available to teachers, a positive number means it
is available for all (See Figure~\ref{fig:AccessPrivileges}). This
naive solution will be changed to a Unix-like permissions system, {\em
  users}, {\em groups} and {\em others} will have {\em read}, {\em
  write}, and {\em visibility} privileges. Every construction will be
associated to a user, who will have all the privileges over the
construction. Teachers will be capable of defining groups, any subset
of the class' students, and to define which kind of privileges
(read/write/visibility) the members of the group will have among them.

\subsection{Technical Issues}
\label{sec:wglTechnicalIssues}

The WGL is a multi-platform, multi-language, multi-tool, synchronous
and asynchronous Web-en\-vi\-ro\-nment. Some of these issues are not new
and also not difficult to solve, others, like the smoothly integration
of tools, raise some important and difficult technical questions.

In this section we will try to list all the questions that should be
addressed by the programmer of such a system and the way the WGL had
addressed, or will address, those questions:

\begin{itemize}
\item multi-platform: this is, in our opinion, a very important
  feature of an educational software because it opens its use to a
  wider audience. To build a multi-platform software it is important
  to choose the right tools and technologies. The WGL uses: {\em
    Apache} (Web server); {\em PHP} (Web script, server side,
  language); {\em MySQL} (DBMS); {\em AJAX} (asynchronous Javascript
  and XML); {\em JavaScript} (Web script, client side, language); {\em
    Java Applets} (Web language). All these tools and programming
  languages are multi-platform;


\item multi-language: another important issue whenever we want to
  build a software system for the global world. In WGL the {\em
    gettext} library~\cite{Drepper2010} for {\em PHP} is used. The
  {\em base\/} system is generic, having as default language the
  English. Adding to the base system a set of translations will
  provide localisation to the system. Currently only the Portuguese
  translation is provided;
\item Web-environment: this issue is an easy to solve issue. From the
  server side the {\em Apache} Web-server is being used, but any other
  Web-server should be able to support WGL. The pages are being built
  using CSS/HTML/PHP (plus other technologies already mentioned). From
  the client-side any Java-aware browser will be able to run the
  environment;
\item synchronous/asynchronous interaction: this is important at two
  different levels: the interaction of the user with the system and
  with each other; the interaction between tools. The first issue is a
  design question, the WGL blended-learning environment allows both a
  synchronous interaction (classroom) and an asynchronous interaction.
  The other question is very important in a multi-tool Web environment
  if we want to provide a fluid, easy to use, system where the
  interaction of the user with the DGS (synchronously) and the
  repository of problems (asynchronously) should be transparent to the
  user. That is, the user should be able to fetch a new construction
  and to load it into the DGS or, on the opposite direction, save the
  DGS construction into the database, and all this without glitches in
  the environment.
  
  In WGL this question is solved using the AJAX asynchronous features,
  allowing loading/saving construction from/to the database without
  interfering with the DGS;
\item integration of tools: this is the most difficult technical issue
  to be dealt with when building a system like WGL. Two main issues
  must be dealt with and solved: the communication between tools and
  its integration in terms of a fluid interaction with the users. Both
  step should be transparent to the users.

  The integration of the DGS (GeoGebra) is done via the DGS API, the
  asynchronous update of the Web-page is done with the help of an AJAX
  call. The DGS API provides the construction in a XML format and this
  text is kept, and recovered when needed, in the database.

  The integration of a GATP is still to be done. We are working on
  the {\sc i2gatp}~\cite{Quaresma08,Quaresma2011}, an extension of the
  {\sc i2g} format~\cite{Santiago2010} to cope with geometric
  conjectures and its proofs. The integration of an GATP, e.g. the
  {\em GCLCprover}~\cite{Janicic06a}, will use this format to
  communicate, via an API, with the DGS and the database.
\end{itemize}

A component still missing is the construction of the student's
interaction model, we foresee no technical difficulties with this task
given the fact that the asynchronous connection with the database is
already done. In a similar way the collaborative features will be
implemented using the asynchronous connections, allowing the exchange
of geometric contents.

\subsection{Installation}
\label{sec:wglInstallation}

That system will be easy to install on a school server (freeing the
teachers of technical issues), or even on the teachers' personal
computers (given more liberty of use to the teacher). It will require:

\begin{itemize}
\item a computer hosting a network server (local or global) capable of
  providing access to the Web pages of the WGL, and a DBMS to provide
  access to the repository of problems and to the students'
  individualised memory. Any Linux/MacOS computer will be capable of
  providing such an environment, it will also be possible to build such
  an environment on top of a MS-Windows system;
\item access to a network (local or global). The school network or
  even, with the help of an wireless access point, a local/classroom
  wireless network provided by the teacher;
\item computers containing a Web browser connected to the network for
  teachers and students access to the system. Nowadays any computer
  is equipped with such a program, if not, installing one is always
  possible. 
\end{itemize}

All this can be built using open source programs so, apart from the
hardware, the WGL system has no other costs, and even the hardware
necessary to support the server and the clients will not represent a
great burden.

\section{A Case Study}
\label{sec:CaseStudy}

We will now try to do a {\em walking tour\/} throughout WGL by means
of an example. We will try to describe the possibilities of the
current implementation and also to point out some of the improvements
already thought out.

For this example we will assume that all the registration of teachers
(by an administrator of WGL) and students (by theirs teachers) is
already done.

The example will be about the notable points of a triangle, e.g. the
incenter, the circuncenter and the orthocenter. The lesson should
have two distinct moments: during the class the introduction of the
incenter and its construction by the students using the WGL; after the
class the other two notable points.

\begin{figure}[hbt!]
  \centering
  \includegraphics[scale=0.3]{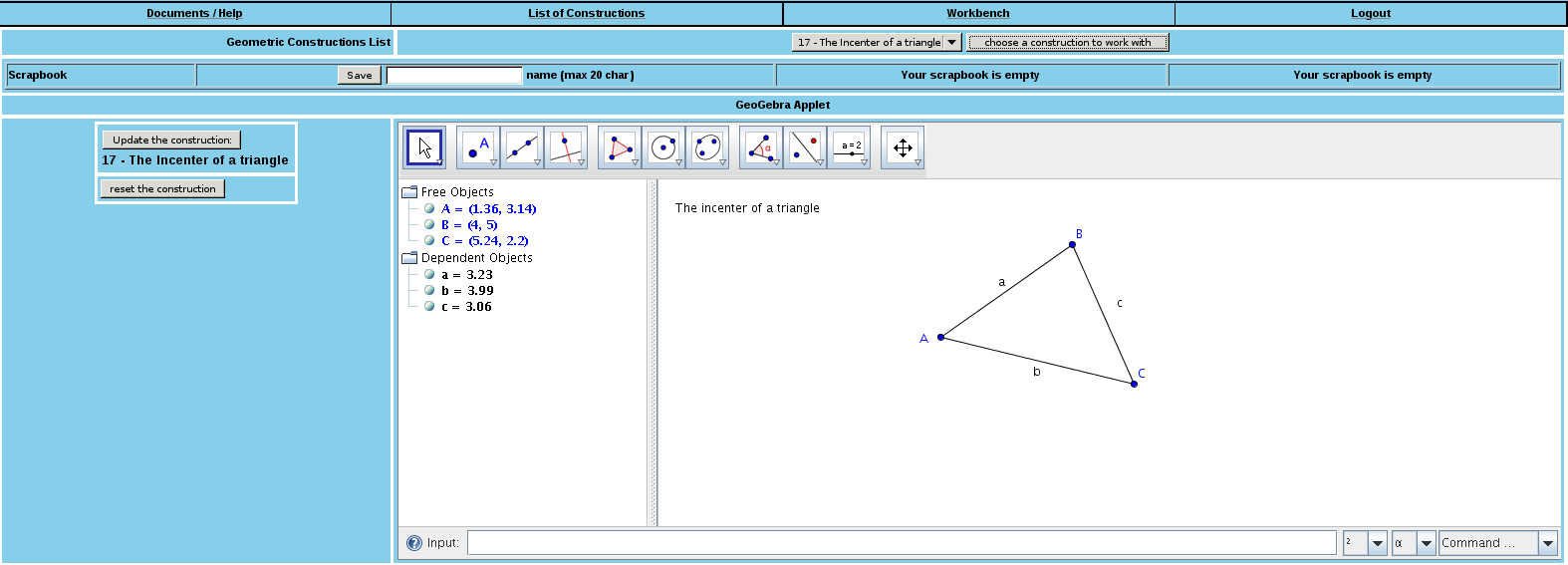}  
  \caption{The Initial Construction}
  \label{fig:InitialConstruction}
\end{figure}

\begin{figure}[hbt!]
  \centering
  \includegraphics[scale=.3]{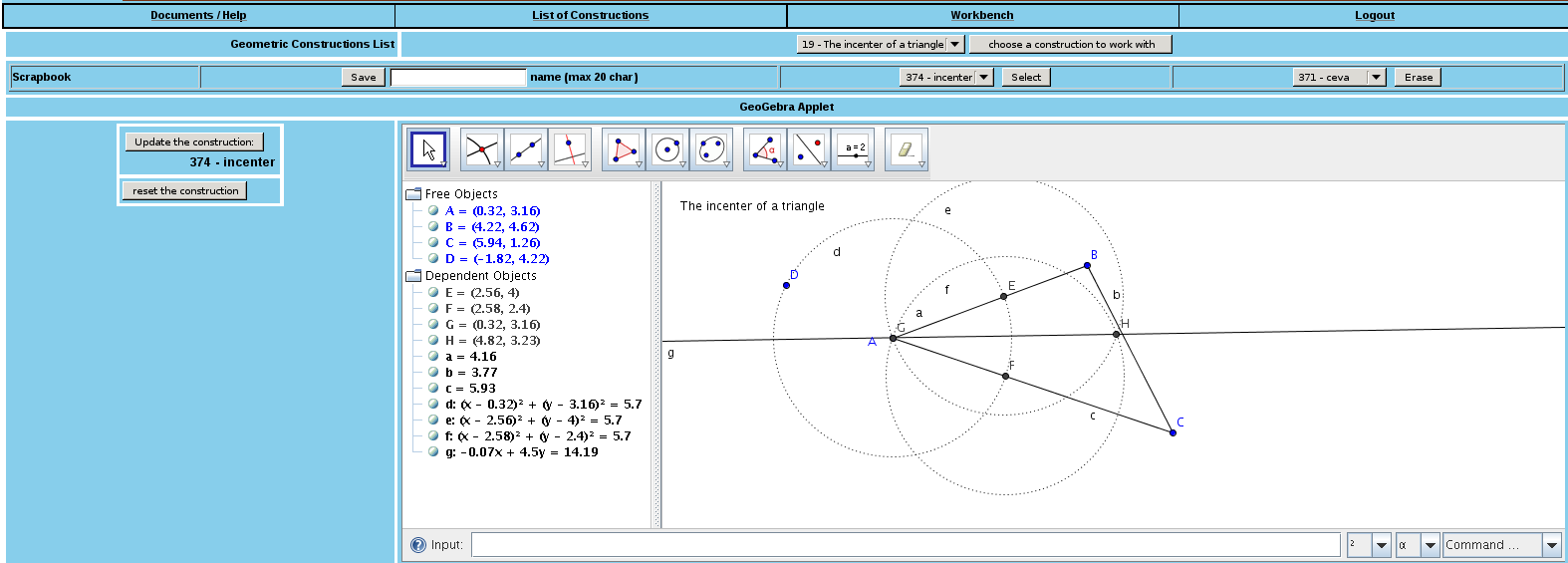}  
  \caption{Angle Bisector}
  \label{fig:AngleBissector}
\end{figure}

\begin{figure}[htb!]
  \centering
  \includegraphics[scale=.3]{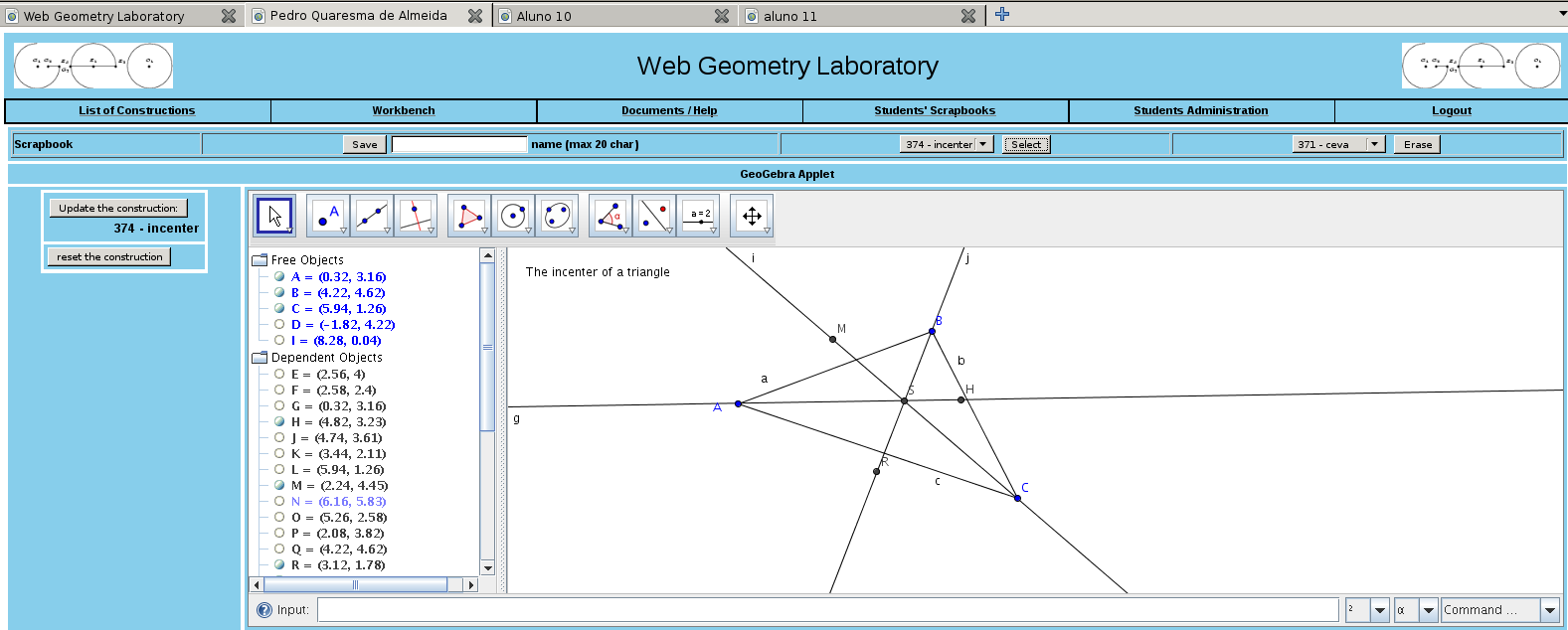}  
  \caption{Incenter \& Students' Tabs}
  \label{fig:StudentsTabs}
\end{figure}

\begin{figure}[htb!]
  \centering
  \includegraphics[scale=.3]{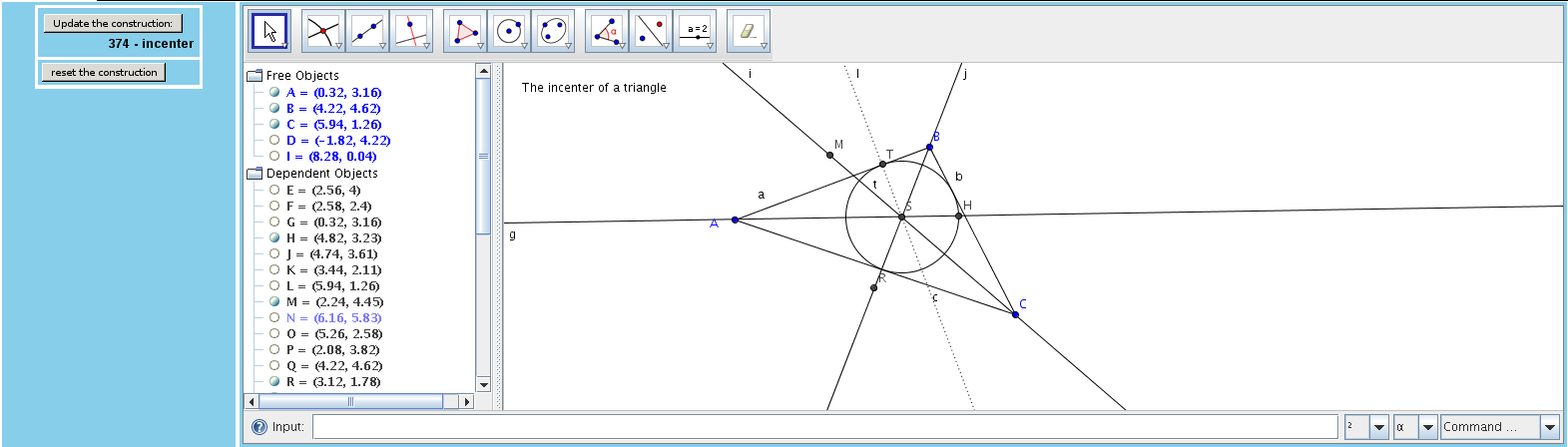}  
  \caption{Incircle}
  \label{fig:Incircle}
\end{figure}

Before the class, during the planning phase, the teacher will use the
WGL to prepare a set of constructions to be delivered during the
class: an initial construction with the triangle
(Figure~\ref{fig:InitialConstruction}); the construction of the first
internal angle bisector (Figure~\ref{fig:AngleBissector}); the
intersection of the three angle bisectors, the incenter,
(Figure~\ref{fig:StudentsTabs}); and the incircle
(Figure~\ref{fig:Incircle}).

\begin{figure}[hbt!]
  \centering
  \includegraphics[scale=.5]{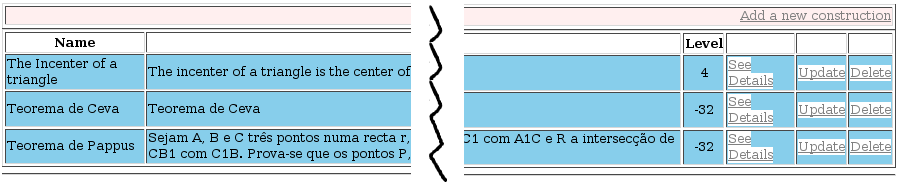}

  \vspace*{1.2em}
  \includegraphics[scale=.5]{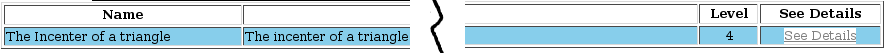}
  
  \caption{Constructions List / Access Privileges}
  \label{fig:AccessPrivileges}
\end{figure}

For every one of these constructions the teacher can decide which ones
will be visible to the students and which ones will be only visible to
himself/herself. In figure~\ref{fig:AccessPrivileges} we can see the
teacher's and the students' view of a given constructions list.

At a certain point during the class, maybe after an initial
introduction of the notable points of a triangle, the teacher will ask
the students to login into the WGL, load the initial construction (see
Figures~\ref{fig:InitialConstruction} and \ref{fig:AccessPrivileges}),
and to begin work on the first task, the construction of the angle
bisector (see Figure~\ref{fig:AngleBissector}).

At this point all the students can access the construction made by the
teacher and they can begin working individually. In a collaborative
perspective the students should be allowed to exchange their work
among the groups and even to be able to make the construction
collaboratively. In the current implementation of WGL only the teacher
has access to all the students work (notice the different tabs with the
students name in Figure~\ref{fig:StudentsTabs}). This should be
greatly improved in a next version of WGL, the implementation of the
users' privileges system should allow extending this to groups of
students working collaboratively.

As said above the teacher has access to all the students'
constructions. At any moment he/she can come to help a student or a
group of students and, when the task is complete, he/she can broadcast
the solution, as the starting point of a new task, e.g. finding the
other bisectors and their intersection point (see
Figure~\ref{fig:StudentsTabs}). The solution broadcast by the teachers
can be a previously prepared (by the teacher) construction or one of
the students' solutions.

The teacher should always stress out the constructive nature of the
geometric figure being made. From a set of free points and using a
well defined set of constructive steps, another set of points is being
constructed. Having finished the incenter construction, the dynamic
possibilities of the DGS could be used to show that this construction
is more generic that it seems at a first glance. This could, and
should be, used to open the door to the rigorous, or even formal,
proof of this result. When the connection between WGL and a GATP
(synthetic or algebraic) will be completed, it will be possible to
write down a geometric conjecture, related to the construction being
made, and then, calling an appropriate GATP, to have access to its
proof, or at least to the proof result.

At the end of the class the teacher will introduce the other notable
points of the triangle, e.g. the circuncenter and the orthocenter,
making visible their initial constructions and setting the tasks to be
performed by the students, with the help of WGL, but outside the
classroom. Again using the users' privileges system this could be done
in a collaborative setting.

All the constructions made by the students, are kept in their personal
scrapbook. This is not, yet, an adaptative system, but it gives
already a certain degree of individualisation, giving to each
student a track record of his/her own work. The teacher has read access to
all the students' scrapbooks. This can be used to monitor their work
and/or to be used to grade the students, evaluating all the work done
by each student, during the classes and also outside the classes.

\section{Conclusions \& Future Work}
\label{sec:ConclusionsFutureWork}

With the integration of these three tools: a DGS, a GATP and a DBMS we
aim to build an adaptative and collaborative blended-learning
environment for geometry.

\paragraph{Conclusions.}
\label{sec:conclusions}
The integration of dynamic geometry tools in the school environment
allows to diversify the study of geometry, enhancing its dynamic side.
One objective is to place the students in front of a ``visual
demonstration'' of several case studies. The ability to perceive
different representations of the same construction is a strategic
point for the students development. The control of geometric
configurations leads to the discovery of new and interesting
properties. The learning process allows to make experiences, develop
strategies, make conjectures, reason and deduce mathematical
properties, and from this to begin using the geometric automated
theorem provers to introduce formal deductive reasoning.

The {\webgeometrylab} will allow teachers to use the DGSs and GATPs in
a more fruitful way. The teacher will be able to prepare a
set of constructions  in advance and to provide them to an entire class easily. It
will also allow the teachers to monitor each student, during the class
and after the classes. At the same time students will have an
individual platform for learning geometry at their own pace.

The goal of this project is to build a dynamic adaptative and
collaborative blended-learning environment for geometry, incorporating
a dynamic geometry software tool, one, or more, geometric automated
theorem provers and a repository of geometric problems, in an
environment where the student has the necessary tools at his/her
disposal for the study of theories and models of geometry. A system
where the student can understand the differences and connections
between these two perspectives and improve their knowledge. A system
in which the student can also be challenged by new problems, giving
the student the opportunity to develop and improve their study in the
area of geometry anywhere, at any time and at their own pace.


\paragraph{Future Work.}
\label{sec:futurework}

The {\webgeometrylab} is a ``work in progress'' project. As a first
task we need to complete a first prototype of a standalone system
capable of being distributed to schools and/or teachers. Such a system
should already include a set of geometric constructions and a course
syllabus to help teachers to organise their work. The integration of a
GATP and/or an interactive theorem proving (ITP) is also a wanted
feature and it is planned as a task to be pursued in parallel with the
described tasks.

The adaptative and collaborative features should be studied to have a
better understanding of how to use them in a geometric context.


\newcommand{\noopsort}[1]{} \newcommand{\singleletter}[1]{#1}


\end{document}